\documentclass[longbibliography,twocolumn,pre,aps,final,showpacs,floatfix]{revtex4-1}
\usepackage[english]{babel}
\usepackage[latin1]{inputenc}
\usepackage{amsmath,bbm}
\usepackage{graphicx}
\usepackage{amsmath}
\usepackage{amsfonts}
\usepackage{euler}
\def\div{{{\rm div}\; }}
\def\dV{{\;d^3 {\bf r}}}
\def\D{{{\bf D} }}
\def\r{{{\bf r} }}
\def\ss{{\rho_f}}
\newcommand{\E}{{\mathbf{E}}}

\renewcommand{\div}{\operatorname{div}}

\newcommand{\toto}{}
\begin{document}
\title{A minimizing principle for the  Poisson-Boltzmann equation}
\author{A.C. Maggs} 

\ifdefined \toto
\affiliation{Laboratoire PCT,
 Gulliver CNRS-ESPCI UMR 7083, 10 rue Vauquelin, 75231 Paris Cedex
  05.}
\fi
\ifdefined \nototo
\institute{
Lab. PCT,
 Gulliver CNRS-ESPCI UMR 7083, 10 rue Vauquelin, 75231 Paris Cedex
  05.
}
\fi
 \begin{abstract}
  The Poisson-Boltzmann equation is often presented via a variational
  formulation based on the electrostatic potential. However, the
  functional has the defect of being non-convex. It can not be used as
  a local minimization principle while coupled to other dynamic
  degrees of freedom. We formulate a convex dual functional which is
  numerically equivalent at its minimum and which is more suited
  to local optimization.
\end{abstract}
\ifdefined \nototo
\fi
\maketitle

\ifdefined \toto
\subsection*{Introduction}
\fi
The Poisson-Boltzmann treatment of free ions is widely used in
simulations with implicit solvents. It successfully describes the mean
field screening properties of ionic solutions. At the most
phenomenological level it is found by balancing two crucial features
of the ionic system: The electrostatic energy coming from Coulomb's
law, plus the ideal entropy of mixing of the ions.  Written in terms
of ion concentrations $c_j$, or the charge concentrations $\rho_j= c_j
q_j$ we find the free energy of a solution from
\begin{align}
  F= \int \dV \dV' \; \rho(\r) \frac{1}{4 \pi \epsilon_0 |\r-\r' | }
  \rho(\r') + \nonumber  \\   k_BT \int  \dV
  \sum_j ( c_j
  \ln{(c_j/c_{j0}) } - c_j) \label{eq:coulomb}
\end{align}
where the total charge density $\rho= \sum_j \rho_j + \ss$.  $\ss$ is
an external fixed charge density -- associated with surfaces or
molecular sources, $c_{j0}$ is the reference density of component
$j$. If we minimize this functional of $c_j$ then we find an effective
free energy for the source field $\ss$, \cite{maccammon}.  The problem
in this formulation is the appearance of the long-ranged Coulomb
interaction \cite{hansen} which renders the evaluation and
minimization less efficient than might be wished.

We start with a few points of notation: We will use the symbol $f$ to
describe a density of free energy of an electrolyte in mean field
theory, $h$ denotes a complex action which occurs in a functional
integral and $u$ is the electrostatic energy of the electric field. We
will freely integrate by parts in our expressions dropping boundary terms.
The transformations that we perform conserve the stationary point of
the mean field solution which is denoted $f_s$.  The statistical
weight of a configuration is then
\begin{equation}
  w = e^{-\beta \int f_s \dV}
\end{equation}
no matter the arguments of the functional $f$. $\ss$ is kept constant
during all variations of the electric and ionic degrees of freedom. We
denote the Legendre transform of a convex function $g(x)$ as
\begin{equation}
  \tilde g(\xi) = {\mathcal L} ( g(x) ) [\xi] = \sup_x ( x \xi -g(x) ) \label{eq:leg}
\end{equation}

In eq.~(\ref{eq:coulomb}) one conventionally decouples the
electrostatic interactions by introducing the potential as an
additional variational field. If we do so then we find that
\begin{equation}
  f=  \rho \phi - \epsilon\frac{(\nabla \phi)^2}{2} +\sum_j  k_BT (c_j \ln
  (c_j/c_{j0}) - c_j) \label{eq:decouple}
\end{equation}
We see at once that we have gained in locality of the formulation; the
other advantage is that it is valid for arbitrary spatial variation in
the dielectric properties, $\epsilon(\r)$.  Unfortunately, the
counterpart is that the resulting free energy is no-longer convex. We
do not have a minimizing principle, rather a stationary
principle. This excludes some of the simplest optimization strategies
that one might want apply -- such as simultaneous annealing of
conformational and electrostatic degrees of freedom.  From
eq.~(\ref{eq:decouple}) one studies the differential
\begin{equation*}
  df =    \sum_j d c_j  ({q_j \phi + k_BT \ln(c_j/c_{j0})} ) +
  d\phi( \rho + \nabla \cdot \epsilon \nabla \phi)
\end{equation*}
We now impose that the coefficient of $dc_j$ is zero so that
\begin{equation}
  q_j \phi+ k_BT \ln(c_j/c_{j0} )=0
\end{equation}
Substituting back into eq.~(\ref{eq:decouple}) we find the standard
form for the Poisson-Boltzmann functional
\cite{briggs,honig,roux,reiner,mccammon}:
\begin{equation}
  f=    \ss \phi - \epsilon \frac{(\nabla \phi)^2}{2} - k_BT \sum_j   c_{j0} e^{-q_j \phi \beta} 
  \ \label{eq:G}
\end{equation}
We take the variations of this functional with respect to $\phi$ to
find:
\begin{equation}
  \ss +   \nabla \cdot \epsilon \nabla \phi +  \sum_j q_j c_{j0} e^{-\beta q_j \phi}=0
\end{equation}
For the symmetric ion system
\begin{align}
  f= & \ss \phi - \epsilon \frac{(\nabla \phi)^2}{2} -2 k_BT c_0
  \cosh{(\beta q
    \phi)} \label{eq:sym}\\
  & \ss+ \nabla \cdot \epsilon \nabla \phi - 2 q c_0 \sinh(\beta q
  \phi)=0 \label{eq:sympb}
\end{align}
Eq.~(\ref{eq:sym}) continues to be awkward numerically since clearly
both the derivative and the $\cosh$ functions are unbounded below;
simple annealing procedures are thus unstable. One has to solve the
partial differential equation eq.~(\ref{eq:sympb}) in practical
applications \cite{gilson, honig2}.

The purpose of this paper is to derive functionals that are equivalent
to those given above which combine the advantages of the convexity of
eq.~(\ref{eq:coulomb}) and the locality of eq.~(\ref{eq:decouple}). We
show how to find functions which are {\it numerically equivalent} to
those widely used in the literature. We consider this {\it absolutely
  crucial} since much time has already been invested in calibration of
potential functions. We thus look for ways of rendering convex known,
accurate functionals. Our main tool in this effort will be the
Legendre transform in a form presented in \cite{courant} as a
reciprocal principle for variational calculations.  In the following
calculations the way that the energy $\epsilon (\nabla \phi)^2/2$ is
transformed into an equivalent form $\D^2/2\epsilon $ will remind the
reader of the transformation of the kinetic energy in a Lagrangian, $m
\dot x^2/2$ into the Hamiltonian equivalent, $p^2/2m$. This is a
standard example of the use of Legendre transforms in classical
mechanics.

We start by summarizing the state of the art in the calculation of
effective free energies for Poisson-Boltzmann free energies, and then
showing in a simple example how the ideas of dynamically constrained
fields can be used to write a convex free energy in terms of the
electric displacement field $\D$.  We then generalize the procedure to
arbitrary formulations of the free energy demonstrating the link to
the theory of Legendre transforms.

\ifdefined \toto
\subsection*{Field theory route to Poisson-Boltzmann}
\fi

In recent years theoretical methods have been introduced to generalize
the application of Poisson-Boltzmann equations to a larger range of
systems, \cite{henri1, henri2}. All these methods are based on the
Hubbard-Stratonovich transformation to break pair interactions into
one-body potentials. They have shown their power in producing
functionals that include the finite volume of ions as well as mobile
dipoles. The Hubbard-Stratonovich transformation produces a complex
action, which can however be studied at its saddle point.  In this
case one reproduces the known Poisson-Boltzmann functionals for
simple situations, and in addition study fluctuations about the
mean field solution in terms of a loop expansion.

A typical example of such a function for both symmetric ions and
dipoles is \cite{dipole, dipoleB, marc}
\begin{align}
f &=   \ss  \phi- \epsilon_0 \frac{(\nabla \phi)^2}{2} \nonumber -\\ & \frac{k_BT}{a^3}  \ln
\left (
1 + 2 \lambda_{ion} \cosh{(\beta q \phi)} + \lambda_{dip} \frac{ \sinh(\beta p_0
 | \nabla \phi| ) }{\beta p_0 |\nabla \phi|}
\right ) 
\label{eq:dipole}
\end{align}
where $\lambda_{ion}$ and $\lambda_{dip}$ are the chemical potentials of the
ions and the dipoles, $a$ a hard core size, $p_0$ a dipole
moment. $\epsilon(\r)= \epsilon_0$, since dielectric effects are
generated dynamically. The logarithmic term in the energy is found from studying the
partition function of a lattice gas.

\ifdefined \toto
\subsection*{Constrained fields plus mixing entropy}
\fi
An alternative, local approach to electrostatics was introduced in
\cite{vincent, joerg,everaers} to avoid the use of Ewald and fast Fourier methods in
electrostatic codes. The ideas were applied to the Poisson-Boltzmann
equation in \cite{maggspb, burkard}. A convex energy is
associated with the Gauss law constraint which is implemented
dynamically. In particular we reproduce the  mean field
Poisson-Boltzmann equations from the functional
\begin{align}
f= &   
\frac{ \D^2}{2\epsilon } +  k_BT \sum_j (c_j \ln{(c_j/c_{j0})} - c_j )
\\
&
{\rm with} \quad  \div 
\D - \rho    =0
\end{align}
In the case of a one-component plasma one  trivially eliminates
the density degree of freedom to find
\begin{align}
f&=  \frac{ \D^2}{2\epsilon } +  k_BT\; s( (\div \D - \ss)/q) \label{eq:cpb} \\
s& (z) = z \ln(z/c_0) -z \label{eq:entropy}
\end{align}
The functional eq.~(\ref{eq:cpb}) is now an unconstrained functional of the field
$\D$. It can be discretized as in \cite{vincent} where the vector
field $\D$ is associated to the links of a cubic lattice, while scalar
quantities such as $(\div \D)$ and $\ss$ are associated with the
vertexes of the lattice. Eq.~(\ref{eq:cpb}) has the advantage of
 being both  convex and local.

\ifdefined \toto
\subsection*{Reciprocity and Legendre transforms}
\fi
We now show how to transform a Poisson-Boltzmann functionals expressed
in terms of electrostatic potentials $\phi$ into those based on the
electric fields,
$\D$. 
Start with the electrostatic  energy density
\begin{equation}
u=  \ss \phi - \epsilon \frac{(\nabla \phi)^2}{2} \label{eq:p}
\end{equation}
As is well known the Poisson equation is found as the stationary
condition of eq.~(\ref{eq:p}).
Introduce now $ \E = -\nabla \phi$ using a Lagrange multiplier $\D$ to
find the constrained functional
\begin{equation}
u=  \ss \phi - \epsilon \frac{\E^2}{2} + \D \cdot (\E + \nabla \phi) \label{eq:grad}
\end{equation}
Integrate by parts and regroup:
\begin{equation}
u =   - \epsilon \frac{\E^2}{2} + \D \cdot \E - \phi(  \div \D - \ss )
\end{equation}
We  now eliminate $\E$ and see the equivalence to
\begin{equation}
u =    \frac{\D^2}{2\epsilon }  - \phi(  \div \D - \ss ) \label{eq:const}
\end{equation}
We now re-interpret the field $\phi$ as a Lagrange multiplier \cite{courant} imposing
Gauss' law and find the starting point of our previous papers \cite{vincent} on
constrained statistical mechanics in electrostatics. 

Let us now generalize this approach to Poisson-Boltzmann functionals
expressed in terms of the potential.
\begin{equation}
f=  \ss \phi - \epsilon \frac{(\nabla \phi)^2}{2} - g(\phi)
\end{equation}
 We will explicitly treat the example of the symmetric electrolyte solution
where $g(\phi) = 2 k_BT c_0 \cosh{(\beta q \phi)}$.  The
transformation of eqs.~(\ref{eq:p}-\ref{eq:const}) still goes through
and we find
\begin{equation}
f =    \frac{\D^2}{2\epsilon }  + \phi( \ss -\div \D ) - g(\phi). \label{eq:div}
\end{equation}
Further variations with
respect to $\phi$ give simply the Legendre transform,
eq.~(\ref{eq:leg}), of the function $g(\phi)$, where the transform
variable is $\xi=(\ss - \div \D)$

To continue the transformation we require the stationary point of
\begin{equation}
\phi\xi - 2 k_BT c_0 \cosh(\beta q \phi) 
\end{equation}
finding  $\phi = (k_BT/q)\sinh^{-1}{(\xi/2qc_0 )}$ and
\begin{equation}
{\mathcal L} (g)[\xi]= \frac{k_BT \xi}{q}  \sinh^{-1}(\xi/2qc_0) - k_BT \sqrt{4c_0^2 +\xi^2/q^2}  \label{eq:inf}
\end{equation}
For symmetric solutions we conclude that the reciprocal functional that we
require is
\begin{equation}
f=  \frac{\D^2}{2\epsilon } + {\mathcal L} (g) \left [ \ss - \div \D \right ]
\end{equation}
This demonstrates the principle result of the present paper. Starting
from a standard, concave functional expressed in terms of the
potential we have found a convex functional of the vector field
$\D$. The critical points of the two functionals are, however,
numerically identical.

When there are no ions within a region -- for instance within a
macromolecule -- the Legendre transform requires some care in its
definition: We take $g(\phi) = \eta \phi^2/2$ with $\eta$
small. Then the Legendre transform is $\tilde g = \xi^2/(2 \eta)$.
Using finite $\eta$ gives a Yukawa or Debye interaction, with
screening of the electrostatics. The limit $\eta \rightarrow 0$
imposes a delta-function constraint on Gauss' law.  Experience with
local Monte Carlo algorithms based on the electric field
\cite{vincent, joerg} implies that relaxation of longitudinal and
transverse degrees of freedom via link, and plaquette updates is the
most efficient manner to sample the above functionals.

\ifdefined \toto
\subsection*{Dipolar systems}
\fi
We now consider the transformation of the functional
eq.~(\ref{eq:dipole}) for small chemical potentials $\lambda_j$, where
we expand the logarithm to lowest order.  In this case we find a
contribution of the form
\begin{equation}
f_{\bf E}= -\epsilon_0 \frac{\E^2}{2} - \bar \lambda_d \frac{\sinh(\beta p_0 |\E|)
}{\beta p_0 |\E|} +\D \cdot \E
\end{equation}
Again if we consider the extremal equation for $\E$ we recognize the
Legendre transform, this time for the electric field, rather than for
the potential. We have here defined $\bar \lambda_j = k_BT \lambda_j/a^3$.

Thus the fully transformed functional in presence of both ions and
dipoles is
\begin{align}
f={\mathcal L}   \left ( \epsilon_0 \frac{\E^2}{2} + 
\bar \lambda_{dip}  \frac {\sinh (\beta p_0|\E |)}  {\beta p_0|\E| }  \right) &  [\D] +
\nonumber \\
2 \bar \lambda_{ion} 
{\mathcal L} \left (\cosh(\beta e \phi) \right) & [ \div \D - \ss ] \label{eq:dipole2}
\end{align}
For small fields, $\E$, we  expand the first line of
eq.~(\ref{eq:dipole2}) and find $\epsilon_0 \frac{\E^2}{2} (1 + \bar \lambda_{dip} \beta^2
p_0^2/3\epsilon_0 ) $.  The modification of the curvature of the function is a
manifestation of the electric susceptibility the dipoles.

\ifdefined \toto
\subsection*{Evaluating transforms}
\fi
In general it is impossible to analytically transform the functions
needed in the dual or reciprocal formulation. In particular the full functional
eq.~(\ref{eq:dipole}) is not a simple sum of terms as assumed in the
derivation of eq.~(\ref{eq:dipole2}). However in a given ionic system
the Legendre transformed functionals are uniform in space. They should
be calculated just once for a given set of chemical potentials. For
the general problem we require the two-dimensional transform with
respect to the variables $(\phi, \E)$. Standard, fast numerical
methods \cite{legendre} are available for performing these transforms,
and the result can be tabulated for a given application. For the
highest accuracy an iterative Newton-Raphson step can be used to
converge the interpolated results to very high accuracy. Forces can be
evaluated as well as energies, for instance in eq.~(\ref{eq:inf}), $d
\tilde g(\xi)/d \xi = \phi$ -- a general property of the transform
\cite{zia}, thus the force on a test particle with charge $e_j$ is just
\begin{equation}
{\bf F}_j = -e_j \nabla \phi = e_j \E
\end{equation}

\ifdefined \toto
\subsection*{Link with Infimal Convolution}
\fi
The Legendre transformation of the entropy function $s(z)$ of
eq.~(\ref{eq:entropy}) is the Boltzmann factor
\begin{equation}
{\mathcal L}(s(z))[\xi] =  c_0 e^\xi
\end{equation}
appearing in eq.~(\ref{eq:G}).  The fact that the effective action
eq.~(\ref{eq:G}) is a sum of exponential functions shows that the
reciprocal action must be linked to {\it infimal convolution} of the
original free energy defined as follows \cite{infimal}:
\begin{equation}
{\mathcal L} ( \tilde g + \tilde d) [x] = \inf_y\;  \left [ g( y ) +
d(x-y) \right ]
\end{equation}
We now show why this is the case for the symmetric electrolyte.

Consider a symmetric electrolyte with unit charges, $q=1$ and
with reference concentrations $c_0=1$.  We impose Gauss' law with a
Lagrange multiplier $\phi$ and study the constrained minimum of the
total ionic entropy
\begin{equation}
t(c_1,c_2)=
s(c_1) + s(c_2) + \phi ( \xi - c_1 + c_2  )
\end{equation}
where $\xi = (\div \D- \ss)$.
The variation with respect to $c_j$ gives
\begin{equation}
\tilde t(\phi) =- \tilde s ( - \phi)  - \tilde s (
\phi) + \phi \xi
\end{equation}
We now study this Legendre transform of the sum of two functions.  For
the symmetric electrolyte we require (from the above definition of the
infimal convolution)
\begin{equation}
 t'(\xi)=\inf_y  \left [  y \ln ( y ) -y   +   (y-\xi) \ln(y-\xi) - (y-\xi) \right ]
\end{equation}
giving $y(y-\xi)=1$, $y= (  \xi + \sqrt{4+\xi^2} )/2$. Thus the action for
the symmetric electrolyte eq.~(\ref{eq:inf}) is found as the infimal
convolution of two ideal entropy functions with the help of the
identity $ \sinh^{-1} (\xi/2) = \ln{ ( ( \xi + \sqrt{4+\xi^2} )/2)} $

\ifdefined \toto
\subsection*{Fluctuations and Fourier transforms}
\fi
We now return to the full field theoretical formulation of the
Poisson-Boltzmann equation after Hubbard-Stratonovich transformation
of the action, now denoted $h$, but before the saddle point evaluation
\cite{henri1,henri2}. As a concrete example we consider the symmetric
electrolyte:
\begin{align}
h =&  \epsilon \frac{(\nabla \phi)^2}{2} - 2 k_BT c_0 \cos{(\beta q \phi})  - i
\ss \phi \\= &  \epsilon \frac{(\nabla \phi)^2}{2} + g(\phi) - i \ss \phi
\end{align}
We now consider transformation in the philosophy of our above reciprocal
formulation, but replacing Lagrange multipliers by complex integral
representations of the delta-function. We do not neglect fluctuations
in the fields and do not make any approximation in the statistical
mechanics of the ionic system. Following very closely the logic
described above we find the following succession of transformations of
the action:
\begin{align}
h=& \epsilon \frac{(\nabla \phi)^2}{2} + g(\phi) - i \ss \phi \\
\rightarrow&  \epsilon  \frac{\E^2}{2} + g(\phi) - i\D \cdot  (\nabla \phi + \E) -
i \ss \phi\\
\rightarrow& \epsilon \frac{\E^2}{2} + g(\phi) - i\D \cdot \E  + i\phi (\div \D -  \ss) \\
\rightarrow&\frac{\D^2}{2\epsilon } + g(\phi)   + i\phi (\div \D -  \ss) 
\end{align}
The only thing that remains is the integral over $\phi$.
This we
recognize as a Fourier transform with variable $\xi=  (\div \D -  \ss)
$.
Thus the action expressed in terms of $\D$ is
\begin{equation}
h = \frac{\D^2}{2\epsilon}  -  \ln \{ {\mathcal F ( e^{-g(\phi)}  ) \} [ \div \D - \ss] }
\end{equation}
with ${\mathcal F}$ the Fourier transform.


This action should be integrated over to find the partition sum
\begin{equation}
Z= \int d \D e^{- \beta \int h \dV}
\end{equation}

When the fluctuations at the saddle point are neglected the Fourier
transform reduces to the Legendre transform as above, and the infimal
convolution is a simple echo of the standard convolution of
statistical weights occurring at the saddle point.  If the Fourier
transform becomes negative then the action becomes complex, or one
must at least sample functions which are non-positive. Again in
general we do not expect to be able to evaluate the Fourier transform
analytically but again tabulation for a specific
problem is possible.

\ifdefined \toto
\subsection*{Conclusions}
\fi
To conclude, we have introduced a duality transformation for
Poisson-Boltzmann functionals which allows us to find a local
minimizing principle for both electrostatic and conformation degrees
of freedom in a simulation.  This opens the perspective of simpler
annealing and dynamic relaxation in molecular simulation, including
local Car-Parrinello evolution of ionic degrees of freedom
\cite{joerg} for complex molecules.  Clearly the interpolation of
source charges to a grid requires control of their self-energy in a
manner which in familiar in molecular dynamics simulations
\cite{holm,darden, shaw}.

The functionals are designed to {\it exactly} conserve the saddle
point.  This is only true in practice if the numerical discretization
is strictly equivalent. The final formulation requires a discretized
divergence for Gauss' law in eq.(\ref{eq:div}), ``${\rm Div}$''. This
divergence operator is adjoint to the ``${\rm -Grad} $'' from
eq.~(\ref{eq:grad}).  Finally the Laplacian in the potential
formulation must obey $\nabla^2 =  {\rm Div\; Grad}$. If this is not
true discretization errors differ between the formulations.

The original and reciprocal functionals have very different forms- as
noted above a weakly constraining energy $\eta \phi^2/2$, with $\eta$
small, is transformed into a steep constraint $\xi^2/(2\eta)$, and
vice-versa. In the future it will be interesting to understand how
this influences mathematical descriptions of fluctuations about the
mean field behaviour. 
As stated in \cite{courant} the potential and field formulations can
be used together to give upper and lower bounds for the mean field
free energy. 


%
\end{document}